\documentclass[aps,twocolumn,showpacs,floatfix,superscriptaddress]{revtex4}
\usepackage{graphicx}
\usepackage{subfigure}
\usepackage{bm}
\usepackage{amsmath}
\usepackage{array}
\usepackage{multirow}
\usepackage{float}
\usepackage{color}
\graphicspath{{figs/}}

\begin{document}

\def\dsum{\displaystyle\sum}
\def\dfrac{\displaystyle\frac}
\def\dint{\displaystyle\int}
\def\bmx{\begin{pmatrix}}
\def\emx{\end{pmatrix}}
\def\Lagr{\mathcal{L}}
\def\E{\mathcal{E}}

\title {Proximity Effects in Topological Insulator Heterostructures\footnote{Project supported by NSF of China (Grant Numbers 91021019, 51074151 and 11034006), National Basic Research Program of China (Grant Numbers 2010CB923401 and 2011CB921801), USDOE (Grant Number DE-FG03-02ER45958), US National Science Foundation (Grant number 0906025), and the BES program of US Department of Energy (Grant Number ER45958).
}}

\author{Xiaoguang Li}
\affiliation{Shenzhen Institutes of Advanced Technology, Chinese Academy of Sciences, Shenzhen, 518055, China}
\affiliation{International Center for Quantum Design of Functional Materials (ICQD)/Hefei National Laboratory for Physical Sciences at the Microscale (HFNL), University of Science and Technology of China, Hefei, Anhui 230026, China}
\author{Gufeng Zhang}
\affiliation{International Center for Quantum Design of Functional Materials (ICQD)/Hefei National Laboratory for Physical Sciences at the Microscale (HFNL), University of Science and Technology of China, Hefei, Anhui 230026, China}
\author{Guangfen Wu}
\affiliation{Shenzhen Institutes of Advanced Technology, Chinese Academy of Sciences, Shenzhen, 518055, China}
\author{Hua Chen}
\affiliation{Department of Physics, University of Texas at Austin, Austin, Texas 78712, USA}
\author{Dimitrie	Culcer} 
\affiliation{International Center for Quantum Design of Functional Materials (ICQD)/Hefei National Laboratory for Physical Sciences at the Microscale (HFNL), University of Science and Technology of China, Hefei, Anhui 230026, China}
\affiliation{School of Physics, The University of New South Wales, Sydney 2052, Australia}
\author{Zhenyu Zhang\footnote{Corresponding author. E-mail: zhangzy@ustc.edu.cn}}
\affiliation{International Center for Quantum Design of Functional Materials (ICQD)/Hefei National Laboratory for Physical Sciences at the Microscale (HFNL), University of Science and Technology of China, Hefei, Anhui 230026, China}

\begin{abstract}
Topological insulators (TIs) are bulk insulators that possess robust helical conducting states along their interfaces with conventional insulators. A tremendous research effort has recently been devoted to TI-based heterostructures, in which conventional proximity effects give rise to a series of exotic physical phenomena. This paper reviews our recent works on the potential existence of topological proximity effects at the interface between a topological insulator and a normal insulator or other topologically trivial systems. Using first-principles approaches, we have established the tunability of the vertical location of the topological helical state via intriguing dual-proximity effects. To further elucidate the control parameters of this effect, we have used the graphene-based heterostructures as prototypical systems to reveal a more complete phase diagram. On the application side of the topological helical states, we have presented a catalysis example, where the topological helical state plays an essential role in facilitating surface reactions by serving as an effective electron bath. These discoveries lay the foundation for accurate manipulation of the real space properties of the topological helical state in TI-based heterostructures and pave the way for realization of the salient functionality of topological insulators in future device applications.
\end{abstract}

\maketitle
\section{Introduction}
While physics plays a vital role in modern technology developments, the milestones of physics are always signified by the application of some fundamental concepts, such as symmetry, quantization, and topology. In condensed matter physics, the central role of the topology has been widely recognized for the quantum Hall effect many years ago.\cite{PhysRevLett.49.405} However, recently, this concept has received the unprecedented attention,\cite{RevModPhys.82.3045, RevModPhys.83.1057} stimulated by the discovery of topological insulators (TIs) theoretically,\cite{PhysRevLett.95.146802, PhysRevLett.95.226801, Bernevig:2006kx} and soon after experimentally.\cite{Konig:2007uq} This nontrivial new class of insulators possesses robust helical conducting states along their interface with conventional insulators. Almost all the fascinate applications of TI is related to this peculiar helical states.

Because of its robustness and Dirac-like dispersion, the helical surface state of TI is not only fundamentally interesting, but is also expected to be practically important. TI-based heterostructures are the natural playgrounds for the realization of the various innovative applications.\cite{Kong:2011tg} Recently, by using different TI-based heterostructures, emergent properties of topological surface states have been observed or predicted.\cite{PhysRevLett.100.096407,PhysRevB.81.241310,PhysRevB.82.195440,PhysRevLett.104.146802,PhysRevB.84.155105,PhysRevB.84.085103,PhysRevLett.107.166801,PhysRevLett.107.056804,PhysRevB.84.201105,Zhang:2012fk,PhysRevLett.109.236804,PhysRevX.2.031004,Culcer:2012oq,Qu:2012qf,2012arXiv1212.1343Z,2013arXiv1304.1275E,PhysRevB.87.161108,PhysRevB.87.085431,Wu:2013fk,PhysRevB.88.024501} For instance, TI/superconductor heterostructures exhibit a superconducting proximity effect offering the possibility of observing Majorana fermions and the realization of non-Abelian topological quantum computing.\cite{PhysRevLett.100.096407, RevModPhys.80.1083} By putting a ferromagnet on a TI, the inverse spin-galvanic effect and giant spin battery effect can be realized.\cite{PhysRevLett.104.146802} Other novel, technologically important properties have also been demonstrated, such as the enhancement of the catalysis process by robust topological surface states in Au-covered TI.\cite{PhysRevLett.107.056804} However, all those applications are faced with one same question that although the helical state is guaranteed by topological arguments, its exact position is not. Actually, in TI-based heterostructures, as two materials touch with each other, the interaction may largely tune the spatial location of the helical state away from the interface.\cite{Zhang:2012fk, Wu:2013fk, 2012arXiv1212.1343Z} These phenomena, we call ``topological proximity effects'', will be reviewed in this paper.

Although the spatial location of the helical states is crucial for TI-related applications, it does not receive enough attention in previous studies of various TI-based heterostructures. In most analytical studies, The helical states are assumed to be local and appear very close to the interface at the TI side.\cite{PhysRevLett.100.096407,Qi:2009nx,PhysRevB.84.201105} Some numerical works have studied the spatial distribution of the helical states, but not in the heterostructures.\cite{1367-2630-12-6-065013,Eremeev:2012xi} Very recently, Black-Schaffer and Balatsky have analyzed the influence of the impurities in the TI surface layer to the helical states.\cite{PhysRevB.85.121103} If we see the contaminated surface layer as a new material, the whole system can then be viewed as a TI-based heterostructure. On the other hand, Zhang {\em et~al.} have investigated spatial location of the helical states in TI-based heterostructures,\cite{Zhang:2012fk} with a Bi$_2$Se$_3$ slab between two Sb$_2$Se$_3$ (or In$_2$Se$_3$) slabs. They found that the topological helical states shift from the edge of nontrivial Bi$_2$Se$_3$ into the bulk of trivial Sb$_2$Se$_3$, even when the thickness of Sb$_2$Se$_3$ extends to around 6 nm, while for an In$_2$Se$_3$ cover, the states have no qualitative change. This phenomenon has been interpreted as a quantum topological phase transition, in which trivial Sb$_2$Se$_3$ transforms into the nontrivial phase and nontrivial Bi$_2$Se$_3$ transforms reversibly into the trivial phase, and meanwhile, the helical states  become Òbulk-likeÓ instead of localizing at the surface. Using both first-principles and tight-binding approaches, we have also studied some TI and conventional insulator (CI) heterostructures.\cite{2012arXiv1212.1343Z,Wu:2013fk} More rich phenomena have been found from our results, while the interpretation is different from that of the ref.\cite{Zhang:2012fk}. But we have also adopted the idea of the quantum topological phase transition, which is consistent with the terminology of the topological proximity effect.

Recent research efforts have revealed that a trivial insulator can be twisted into a topological state by manipulating the spin-orbit interaction,\cite{Xu:2011ve, PhysRevX.1.021001} the crystal lattice,\cite{Chadov:2010mq, PhysRevLett.105.096404, Lin:2010wm} or with the application of an external electric field,\cite{Kim:2012ly} driving the system through a topological phase transition. Motivated by these works and based on an assumption that the helical states marks the boundary between the topologically trivial and nontrivial phases, it is quite natural to conclude that the shift of the helical states in the heterostructures indicates the occurrence of the topological phase transition due to the spatial proximity of different pieces of the heterostructures. Since the topological phase is a nonlocal and global property of the system,\cite{qi:33} the topological phase transition does not involve symmetry breaking associated with a local order parameter.  We term the observed effects as the topological proximity effect to distinguish it from the traditional proximity effects, which typically invoke some symmetry breaking processes measured by the corresponding local order parameters.\cite{PhysRevLett.98.237201,PhysRevLett.100.096407,PhysRevLett.101.066802}

In this paper, we review some recent studies of TI-based heterostructures. The examples presented are mainly taken from our own results for TI/CI heterostructures. The paper is organized as follows. After a brief review of current status of TI-based heterostructures, we show our first principle calculation for a set of TI-based heterostructures and the peculiar ``topological proximity effect'' in Section 2. In section 3, we use the graphene-based TI/CI heterostructure to exhibit a more complete phase diagram for the topological proximity effect. The attempt to control the position of helical state has some clear advantage, which is emphasized with a potential catalytic application in Section 4. In the end, we close the paper with conclusions and outlook in Section 5.

\section{Conventional/Topological Insulator Heterostructures}

The topological helical state arises from the strong intrinsic spin-orbit coupling (SOC) that drives the inversion of the valence and conduction bands to achieve a topologically nontrivial phase in the otherwise conventional insulator (CI).\cite{Bernevig:2006kx, Konig:2007uq} The requirement of the strong SOC and band-inversion process suggests that heavy-element, small-band-gap semiconductors are the most promising candidate materials to reach the topological insulators (TIs). 

\begin{figure}[tbh]
	\includegraphics[width = 3.4 in]{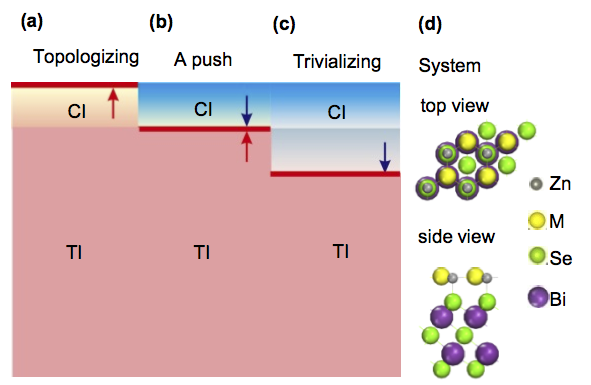}
	\caption{Schematic illustration on tuning the vertical location of the topological surface states (helical state) as a topological insulator (TI) is covered with a layer of conventional insulator (CI). (a) helical state floating to the top of the CI. (b) Staying put at the CI/TI interface. (c) Diving into the TI. (d) The atomic structure of Zn$M$/Bi$_2$Se$_3$ ($M$ = S, Se, Te). The red lines denote the helical state; the arrows indicate the resulting directions of the topological phase transition.}
	\label{fig:gf1}
\end{figure}

Based on the above understanding, we have designed the TI-based heterostructure formed by a single stoichiometric Zn$M$ ($M$ = S, Se, and Te)  layer deposited onto a TI substrate (Bi$_2$Se$_3$ or Bi$_2$Te$_3$). A bulk Zn$M$ has a zincblende structure with its (111) surface matching well with the (0001) surface of the chosen substrate (Fig.~\ref{fig:gf1}d). More importantly, this family of CI span a considerable range of the key parameters such as the SOC strength and band gap, which are expert to conduct different electronic behaviors under the coupling with the proximate TI. Indeed, by using first-principles approaches, we have demonstrated that for the different CI overlayer, the helical state can float to the top of the CI film, or stay put at the CI/TI interface, or be pushed down deeper into the otherwise structurally homogeneous TI substrate (Figs.~\ref{fig:gf1}a-c). These contrasting behaviors imply a rich variety of possible quantum phase transitions in the hybrid systems, dictated by key material-specific properties of the CI. 

\begin{figure*}[tbh]
	\includegraphics[width = 6 in]{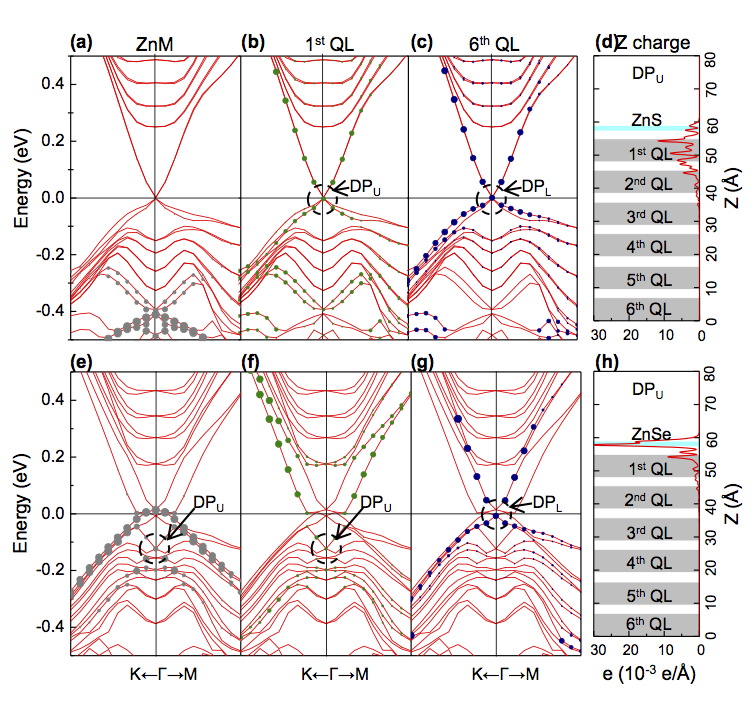}
	\caption{Band structures of ZnS/Bi$_2$Se$_3$ (upper row) and ZnSe/Bi$_2$Se$_3$ (lower row) along the K-$\Gamma$-M direction. The dots indicate the electronic bands contributed by the CI (a) and (e), the 1st QL of the TI (b) and (f), and the 6th QL (c) and (g), respectively; the sizes and colors of the dots also indicate different spectral weights and contributions from different atoms, respectively. (d) and (h) show the charge density distribution of the upper-surface helical state at the $\Gamma$ point marked by the circle and indicated by DP$_\text{U}$. The DP$_\text{U/L}$ stands for the Dirac point at the upper/lower surface. The grey and cyan bars denote the locations of the different QLs and the CI, respectively.}
	\label{fig:gf2}
\end{figure*}

To see the role of the SOC strength and band gap of CI in the heterostructures, we have compared the band structures of  the ZnS/Bi$_2$Se$_3$ and ZnSe/Bi$_2$Se$_3$ systems in Fig.~\ref{fig:gf2}. We note that bulk ZnSe has a much larger SOC and smaller band gap than that of ZnS, in the sense that ZnSe is ÔÔmore proximateÕÕ to a TI material property-wise. This difference may explain the following contrasting proximity effect in the heterostructures. The atom-specific character of each band is indicated by the dots superposed onto the band structure. In this study, the thickness of the TI substrate is chosen to be $6$ quintuple layers (QLs) to insure a negligible interaction between up and down surfaces. From Figs.~\ref{fig:gf2}a and e, we see that ZnS carries negligible electron weight for the helical state, whereas the helical state have a considerable weight on the ZnSe. This difference is further confirmed by the real-space density of states of the upper-surface helical state at the Dirac point (Figs.~\ref{fig:gf2}d and h), where the different spatial locations of the helical states are explicitly shown. From the aspect of topological proximity effect, the ZnS layer remains as a topologically trivial phase, leaving the helical state at the Bi$_2$Se$_3$ surface. On the other hand, the ZnSe layer undergo a topological phase transition making the heterostructure as an entirety into an expanded nontrivial TI phase, and the whole system thus possesses the helical state at the global boundary between the topologically trivial vacuum and the topologically nontrivial heterostructure.

\begin{figure*}[tbh]
	\includegraphics[width = 6 in]{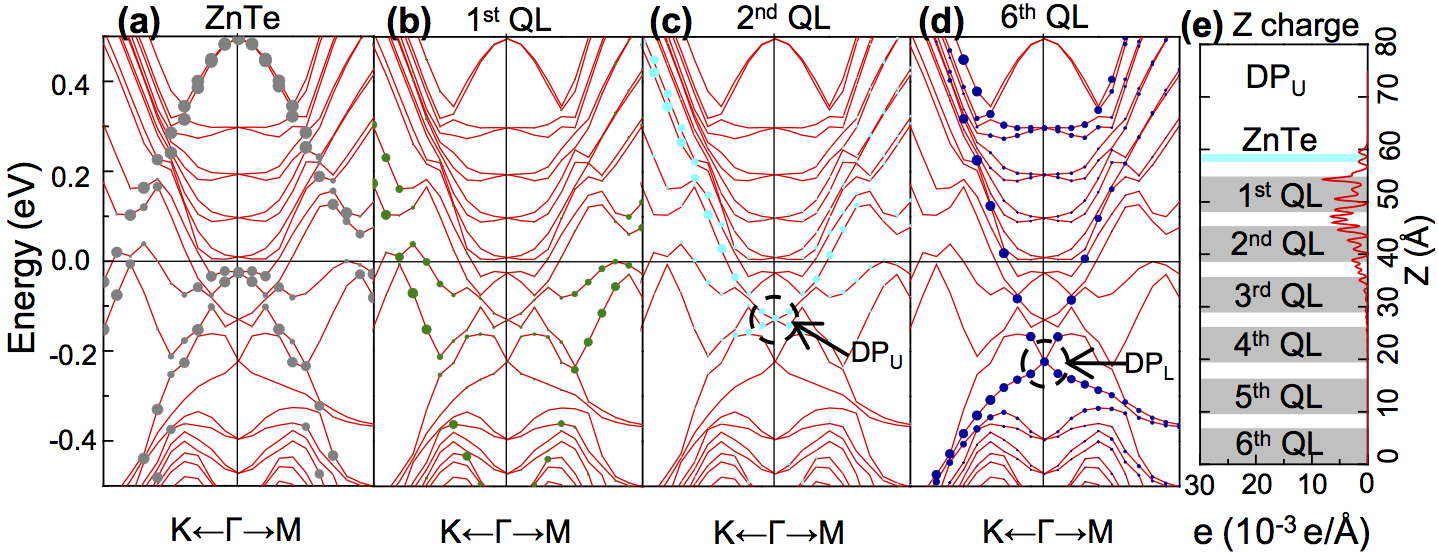}
	\caption{Band structures of ZnTe/Bi2Se3 along the KÐ$\Gamma$ÐM direction. The dots indicate the electronic bands contributed by the CI (a), the 1st QL of the TI (b),the 2nd QL of the TI (c), and the 6th QL (d), respectively; the sizes and colors of the dots also indicate different spectral weights and contributions from different atoms, respectively. (e) shows the charge density distribution of the upper-surface helical state at the $\Gamma$ point marked by the circle and indicated by DP$_\text{U}$.  All other symbols are the same as in Fig.~\ref{fig:gf2}.}
	\label{fig:gf3}
\end{figure*}

With the speculation of the properties-wise proximity in mind, we next look into the system of ZnTe/Bi$_2$Se$_3$. Since ZnTe has an even larger SOC and a smaller bulk band gap than that of ZnSe, the first intuition is to see an even more pronounced relocation of the helical state to the top of the ZnTe film. However, as a counterintuitive surprise, the upper-surface helical state now has a reverse shift toward the inner of Bi$_2$Se$_3$, with its peaked density located on the top of the 2nd QL (Figs.~\ref{fig:gf3}e).

To reveal the physical origin of this unexpected reverse-proximity effect, we first note that, from Figs. 3(a), both the bulk bands of Bi$_2$Se$_3$ and the helical state have a more noticeable downward shift in energy compared to the case of ZnSe/Bi$_2$Se$_3$, while the ZnTe film becomes strongly p-doped, indicating the most pronounced charge transfer from the CI overlayer to the TI substrate among the three cases. This behavior can be understood from the much smaller work function of the ZnTe film in comparison to that of ZnSe. Because of the significant charge transfer, the binding energy between the ZnTe film and the Bi$_2$Se$_3$ substrate is much larger than the other two systems. Since the QLs of Bi$_2$Se$_3$ are mutually coupled through weak van der Waals-like interactions, such a strong coupling between the ZnTe film and the 1st QL will compete with and weaken the interaction between the 1st and 2nd QL. As a consequence, the ZnTe film forces the 1st QL of the TI to be electronically partially decoupled from the remaining QLs. Furthermore, because one QL of Bi$_2$Se$_3$ does not have gapless helical state, the upper-surface helical state will naturally be relocated to the top of the 2nd QL. In other words, the CI has prevailed by topologically ``trivializing'' the 1st QL of the TI via the reverse-proximity effect.

By using the first-principles calculation, a more complete comparative study has also been performed to confirm the role of the interlayer coupling for the proximity effects. Through careful comparison and analysis for the band structures, we have obtained a qualitative understanding about the correlation between the rich proximity effects and some key parameters in the heterostructures. In the next section, we illustrate a more complete phase diagram, by using graphene-based heterostructure as a prototypical system and continuously tuning the parameters, such as the SOC strength, band gap, and the interface coupling between two materials.

\section{A 2D Prototypical System: Topological Graphene Nanoribbon Heterostructures}

In this section, we introduce our studies of the graphene nanoribbon heterostructures to exhibit a more complete picture for the topological proximity effect. Although the intrinsic SOC in graphene is commonly known to be very weak,\cite{PhysRevB.75.041401} as a model system, graphene is the first material predicted to be a 2D TI and indicates the design principle of the TI materials.\cite{PhysRevLett.95.226801,PhysRevLett.95.146802} Therefore, we could expect that the obtained features of topological proximity effects are also suitable for other TI-based systems.

\begin{figure}[tbh]
	\includegraphics[width = 3.4 in]{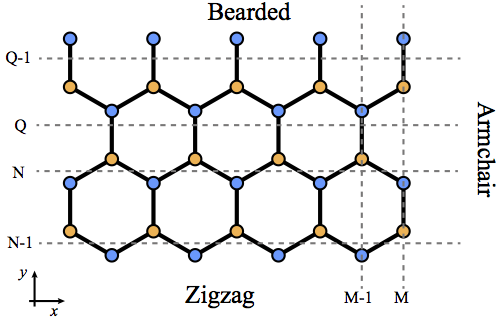}
	\caption{The graphene ribbon geometry with zigzag, bearded and armchair edges. The
dashed lines in zigzag, armchair, and bearded edge labeled with $N$, $M$, and $Q$ indicate $N^{th}$, $M^{th}$ and $Q^{th}$ unit cell in $y$, $x$, $y$ direction, respectively. The width of a ribbon is measured by unit cell. Ribbons of zigzag and beared edge have infinite length in $x$ direction, while for armchair edge in $y$ direction. }
	\label{fig:zgf1}
\end{figure}

\begin{figure*}[tbh]
	\includegraphics[width = 7 in]{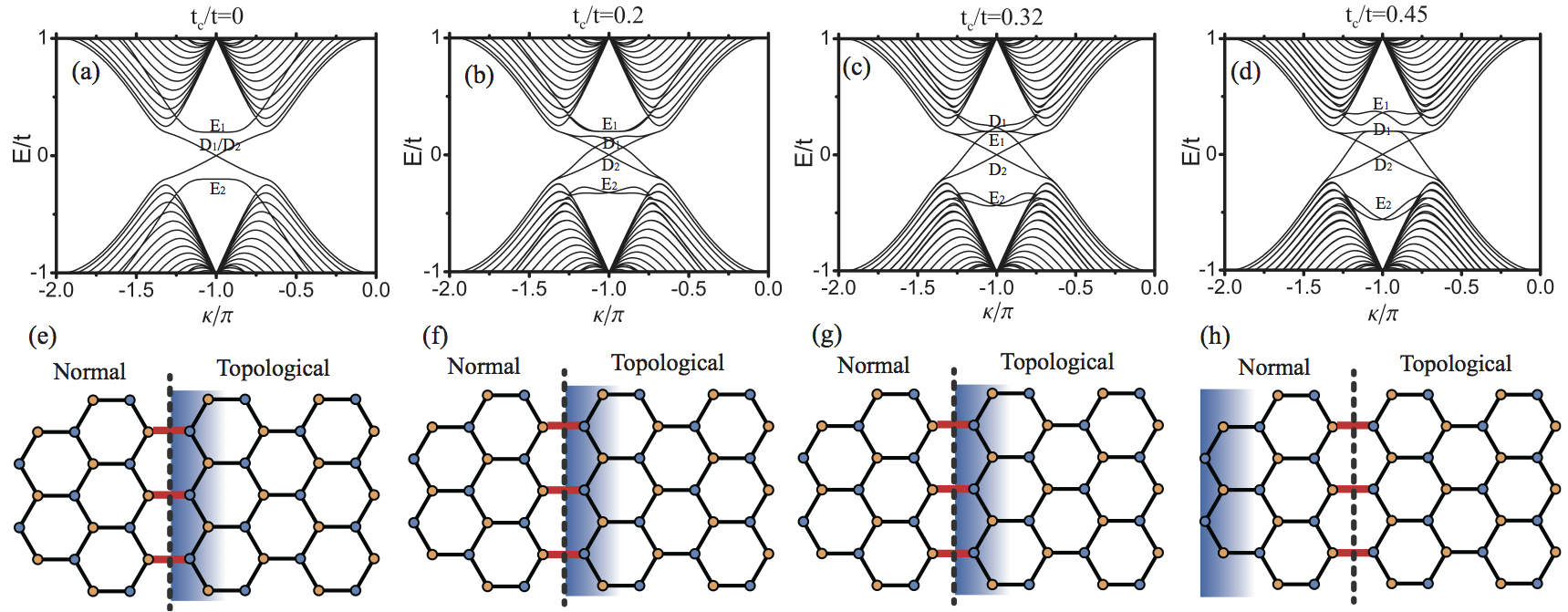}
	\caption{Band structures of zigzag edge GNR heterostructures ($W_n=3$, $W_t=30$) for different tunnel coupling strengths (a-d), and corresponding spatial locations of higher energy Dirac points (denoted as $D_1$ in (a-d)) are shown by blue areas in (e-h). The interface tunnel coupling $t_c/t=0$ in (a)(e), $t_c/t=0.2$ in (b)(f), $t_c/t=0.32$ in (c)(g), $t_c/t=0.45$ in (d)(h). For CI ribbons, $V_g=0.2\,t$ and $\lambda_{SO}/t=0$; for TI ribbons, $V_{g}=0$ and $\lambda_{SO}/t=0.03$. $E_1$ and $E_2$: the bands of the trivial edge states originated from the CI ribbon. $D_2$: the other Dirac point whose spatial location is far away from the interface. Dashed lines in (e-h) indicate the interface of the heterostructures, and the red bonds show the interface tunnel coupling ($t_c$).}
	\label{fig:zgf2}
\end{figure*}

Graphene is a simple two-dimensional carbon system with the honeycomb lattice structure consisting of two atoms in a unit cell.\cite{neto:109} We evaluate the band structure of graphene within the tight-binding Kane-Mele model \cite{PhysRevLett.95.226801,PhysRevLett.95.146802} as follows,
\begin{equation}
\begin{split}
H = &t\sum_{\langle ij\rangle}c_i^{\dag}c_j+\sum_{i\in{a,b}} V_i c_i^{\dag}c_i \\
&+i\,\lambda_{SO}\sum_{\langle\langle ij\rangle \rangle}c^{\dag}_{i} \bm{\sigma}\cdot (\bm{d_{kj}}\times \bm{d_{ik}})\,c_{j},
\label{Eq:KM}
\end{split}
\end{equation}
where $c_i^{\dag}$($c_i$) is the electron creation (annihilation) operator on site $i$; $t$ is the nearest-neighbor hopping strength; $V_{a(b)}$ is the on-site energy for the $A(B)$ sublattice; $\lambda_{SO}$ is second-nearest-neighbor coupling strength determined by the intrinsic SOC; $\bm{\sigma}$ is the Pauli matrix vector; $i$ and $j$ are two next-nearest neighbor sites, $k$ is their unique common nearest neighbor; The vector $\bm{d_{ik}}$ points from $k$ to $i$.

The band structure of graphene can be qualitatively changed by tuning the coupling parameters in Eq.~(\ref{Eq:KM}). When $V_a=-V_b=V_g/2$ and $\lambda_{SO} = 0$, we can obtain a trivial insulator with the band gap equal to $V_g$. On the other hand, when $V_g=0$, we will get a nontrivial insulator with the band gap as $6 \sqrt3 \lambda_{SO}$. Essentially, the competing between $V_g=0$ and $\lambda_{SO}$ determines the topological phase of the graphene. Therefore, by joining two graphene nanoribbons (GNRs) with suitable $V_g$ and $\lambda_{SO}$, we can construct the graphene-based TI/CI heterostructures as in our first-principles studies.

Comparing with our first-principles studies, we have found one more potential location of the helical state in the graphene-based TI/CI heterostructures. In addition, besides the dependence of the SOC and interface coupling, the topological proximity effect also relies on the interface orientations, which could be zigzag, bearded and armchair for a graphene nanoribbon as shown in Fig.~\ref{fig:zgf1}. For zigzag and bearded ribbons, the helical state can be tuned to be either at the interface or the outer edge of the CI ribbon. For armchair ribbons, the potential location of the helical state is further enriched to be at the edge of or within the CI ribbon, at the interface, or diving into the TI ribbon. 

In Fig.~\ref{fig:zgf2}, we exhibit a series of  band structures of the zigzag edge GNR heterostructures, which have the fixed parameters $V_g$ and $\lambda_{SO}$ in both the ribbons, while the different tunnel coupling strength$t_c$ between them. Figure~\ref{fig:zgf2}(a) displays the $t_c = 0$ case, namely, the TI and CI ribbons are decoupled with each other. The band structure is thus the superposition of the TI and CI ribbons. The two degenerate Dirac points ($D_1$ and $D_2$) correspond to the helical states located at the two edges of the TI ribbon, while $E_1$ and $E_2$ bands are two edge states of the CI ribbon. As we gradually increase the tunnel coupling $t_c$, both the energies and spatial locations of these edge states will change accordingly. The phase transition happens between the $t_c$ of Fig.~\ref{fig:zgf2}b and c, indicated by the spatial location switch of $D_1$ and $E_1$. Since the states $E_2$ and $D_1$ locating at the TI/Cl interface are spatially more close to each other, we see from Fig.~\ref{fig:zgf2}a and b that they are more sensitive to $t_c$ before the phase transition. On the other hand, further increasing $t_c$ after the phase transition will not strongly affect the energy of $D_1$, but will keep changing the energy of $E_1$ and $E_2$, implying that these two states are spatially more close now. For very large $t_c$, $E_1$ and $E_2$ mix with bulk states, and we can see only two helical states locate separately at two edges of the whole heterostructures. The shift of topological edge states from the interface to the outer edge of the CI ribbon indicates that the heterostructure as an entirety becomes an expanded TI via the topological proximity effect.

\begin{figure}[h]
	\includegraphics[width = 3.2 in]{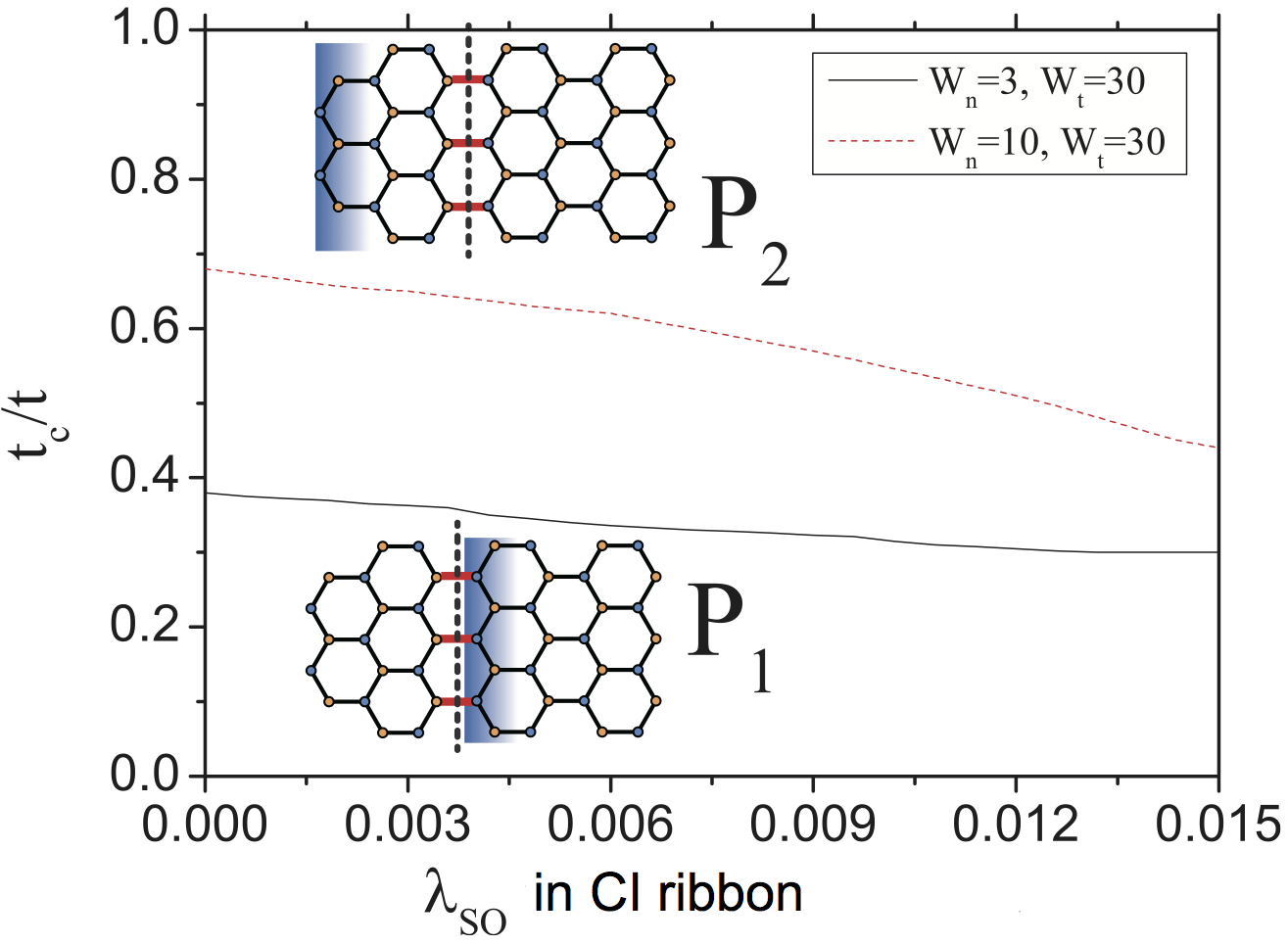}
	\caption{Phase diagram for the zigzag GNR heterostructures with different CI ribbon widths, spanned by the tunnel coupling ($t_c$) and SOC in the CI ribbon. The black solid (red dashed) line indicates the boundary of $P_1$ and $P_2$ phases with $W_n=3$, $W_t=30$ ($W_n=10$, $W_t=30$). Insets are the illustrations of the spatial locations of the helical states. $P_1$ phase: the helical state is located at the interface. $P_2$ phase: the helical state is located at the outer edge of the CI ribbon.}
	\label{fig:zgf3}
\end{figure}

To confirm the speculation in our first-principles studies, we now add a small SOC on the CI ribbon to make it properties-wise more proximate to the nontrivial phase. Under the influence of both the tunnel coupling and SOC, the system still has two phases, indicated by $P_1$ and $P_2$ in Fig.~\ref{fig:zgf3}. As we would expect, the larger $\lambda_{SO}$ makes the topological proximity effect happen more easily, namely, at smaller tunnel coupling $t_c$. Furthermore, we discuss how the width of CI ribbon affects the phase transition. In Fig.~\ref{fig:zgf3}, we see that for the same SOC in CI ribbon, the wider one requires larger tunnel coupling $t_c$ to induce the transition from $P_1$ to $P_2$. This is because the phase transition essentially needs the coupling between the states at the interface and outer edge of the CI ribbon, and this coupling becomes weaker as the width of ribbon increases. In a limiting case where the CI ribbon has infinite width, topological proximity effect will not happen for any finite tunnel coupling $t_c$.

\begin{figure}[h]
	\includegraphics[width = 3.0 in]{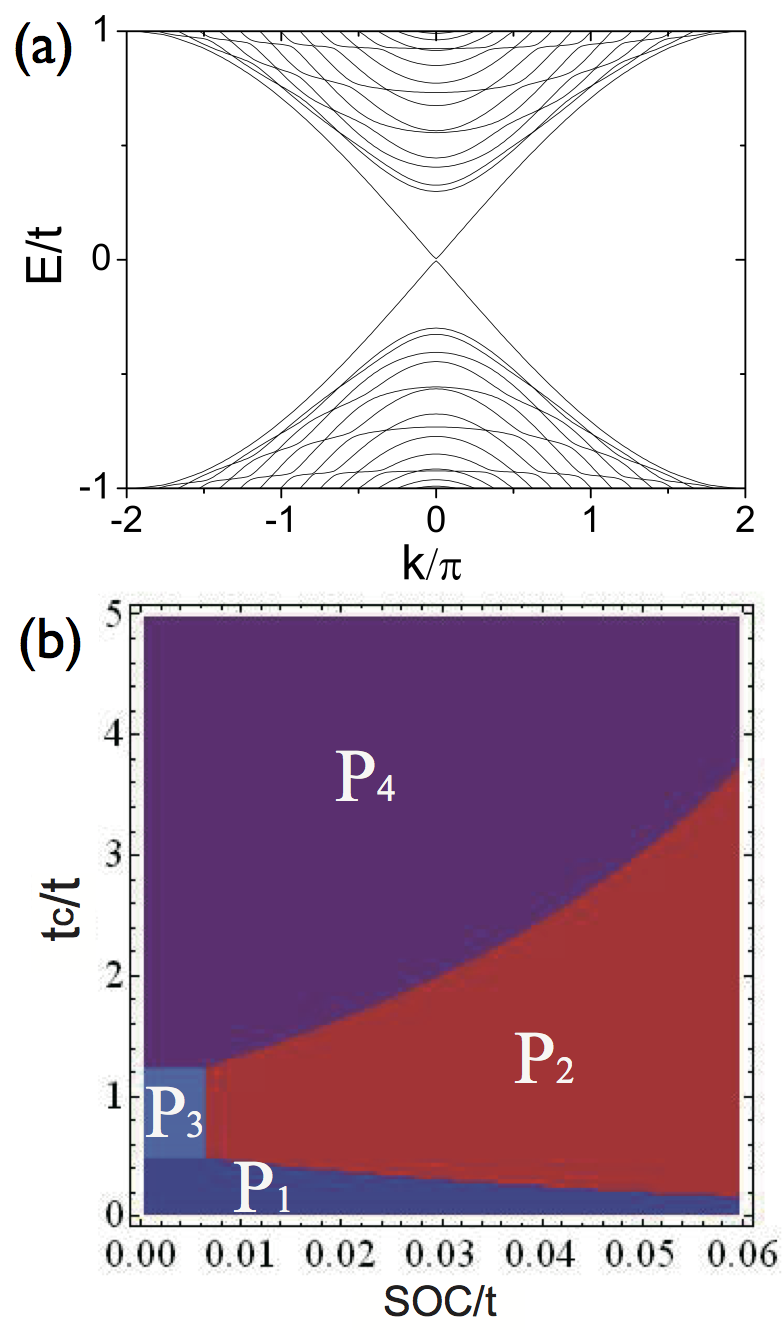}
	\caption{(a) Band structure of an armchair GNR heterostructure whose width is $W_t=30$. The Dirac point is at the $\Gamma$ point. (b) Phase diagram for the armchair GNR heterostructures with a CI ribbon with width $W_n=8$ and a TI ribbon with width $W_t=60$. The diagram is spanned by the tunnel coupling($t_c/t$) and SOC in the CI ribbon. $P_1$ and $P_2$ phases are defined the same as in Fig.~\ref{fig:zgf3}. In $P_3$ phase, the helical state remains in the CI ribbon, whereas in $P_4$ phase the helical state has moved into the TI ribbon by one unit cell.}
	\label{fig:zgf4}
\end{figure}

We now turn to the phase diagram of the armchair edge GNR heterostructures in Fig.~\ref{fig:zgf4}b. In contrast with the zigzag heterostructure, the armchair system shows two additional phases in the strong tunnel coupling regime: $P_3$, where the density of helical state peaks in the bulk of CI ribbon, and $P_4$, where the helical state is relocated one unit cell back inside the topological GNR. With relatively small SOC in the CI ribbon, the location of the helical state shifts from $P_1$ through $P_3$ to $P_4$ as the coupling increases, while for a moderate SOC, the helical state can move to the outer edge of the CI ribbon through $P_2$ instead of $P_3$ during the evolution. In the limiting case of the strong SOC, the original CI ribbon becomes topologically nontrivial.  The appearance of the $P_4$ phase is clearly consistent with our first-principles works, where since only a single layer of CI is considered, $P_3$ phase cannot be verified. We note that the first-principles work by Zhang {\em et~al.} has actually obtained the similar $P_4$ phase, but their interpretation is different from our topological proximity pictures. The movement of the helical state to different position is a manifestation of the complexity of the topological proximity effect. The tunnel coupling $t_c$ plays an interesting role that it effectively propagates the effect of SOC from the TI to CI inducing the phase transition, and meanwhile, it strongly affects the surface layer of the TI making it decoupled from the TI bulk. The competing between these two aspect gives the evolution route from $P_1 \rightarrow P_3 \rightarrow P_4$. When the SOC in CI is large enough that the CI can be entirely conducted to nontrivial phase, $P_2$ will instead $P_3$ phase in the above transition process.

To explain the dependence of the topological proximity effect on the lattice orientation. We compare the band structures of zigzag and armchair edge GNRs around their Dirac points at M (Fig.~\ref{fig:zgf2}) and $\Gamma$ (Fig.~\ref{fig:zgf4}a) points in k space of the zigzag and armchair heterostructures, respectively. For the zigzag system, the M($\Gamma$) points in k space are energetically far away from the bulk states. Correspondingly, in real space, the states at M($\Gamma$) points are edge states that cannot mix with the bulk states, thereby ruling out the existence of the $P_3$ and $P_4$ phases. However, for the armchair system, the $\Gamma$ point represents the helical state and the energy of the $\Gamma$ point is close to the bulk bands. As a consequence, it is possible for the helical state to interact with the bulk states and move into the bulk. We note that most known TIs have similar band structures as armchair GNR, namely, the Dirac points induced by SOC are located at the same $k$ point as the bulk band gap (Fig.~\ref{fig:zgf4}(a)).  

By now we have shown a complete phase diagram of the topological proximity effect by using the 2D graphene-based heterostructures. Also the obtained phenomena are consistent with the three-dimensional (3D) TI/CI system investigated by using the first-principle calculations. In the next section, we present a potential catalytic application of the TI helical states, provided that we can well control their spatial locations. 

\section{A potential catalytic application: Gold Thin Films on Topological Insulator Substrates}

Many physical and chemical processes happening at the material surface are due to the presence of the surface states. However, since they arise as a result of the different bonding environment at the surface from the bulk, normal surface states are easily destroyed by local modifications at the surfaces, e.g., presence of impurities, surface defects, surface reconstruction, or a change in the surface termination or orientation. In contrast, the topological helical state of 3D TI are robust compared to the conventional surface states.\cite{PhysRevLett.98.106803, Hsieh:2008vn, PhysRevLett.103.266803} The helical state protected by time-reversal symmetry is insensitive to the structural details of the surface.\cite{Hsieh:2009lp, Chen:2009ta} It thus provides a perfect platform for investigating the catalytic role of surface states in less constrained environments.

In this section, we introduce a potential catalytic application of the topological insulator heterostructures, by using CO oxidation on gold-covered Bi$_2$Se$_3$ as a prototype example. In this example, the nontrivial helical states resulting from the topological proximity effect, are found to significantly enhance the binding energy of both CO and O$_2$ molecules by serving as an electron bath, and therefore facilitate the oxidation process.

\begin{figure}[h]
	\includegraphics[width = 3.4 in]{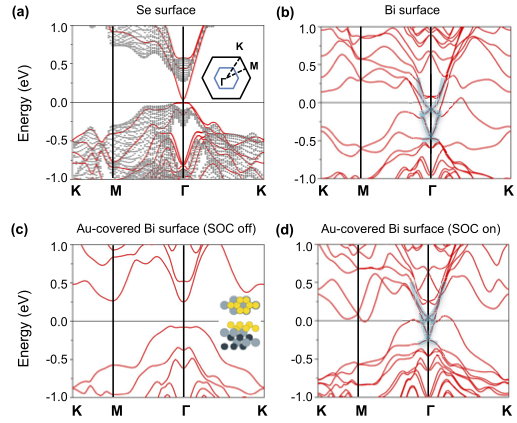}
	\caption{(a) Band structure of a 3 QL Bi$_2$Se$_3$ film, where the shaded area is the bulk band structure projected to the 2D Brillouin zone. The inset shows the shape of the 2D Brillouin zone for different surface unit cells (Black solid lines-1$\times$1; blue solid lines-2$\times$2). (b) Band structure of a Bi-terminated Bi$_2$Se$_3$ film. (c) and (d) Band structures of 2 ML Au deposited on the Bi-terminated surface without and with SOC, respectively. In (b) and (d) the helical state are highlighted by the transparent blue lines. The inset in (c) shows the top and side views of the structure (only top 4 atomic layers of Bi$_2$Se$_3$ are shown). Yellow balls-Au; light blue balls-Bi; dark blue balls-Se.}
	\label{fig:hua1}
\end{figure}

Crystalline Bi$_2$Se$_3$ has rhombohedral structure and its unit cell is composed of 3 weakly coupled QL, each of which comprises of 5 alternating Bi and Se layers in a sequence Se-Bi-Se-Bi-Se. The naturally cleaved surface is therefore the Se surface. In Fig.~\ref{fig:hua1}a we plot the band structure around $E_F$ of a naturally cleaved 3 QL Bi$_2$Se$_3$ film. The surface states are signified by the ``Dirac-cone'' like band structure. The small gap ($\sim$50 meV) opened at the Dirac point is due to the coupling between the two degenerate helical states on the two surfaces of the TI film.\cite{Zhang:2010bu}

We use the Bi-terminated surface of Bi$_2$Se$_3$, to which Au binds more strongly than the naturally cleaved Se surface, to support Au atoms without formation of 3D Au clusters. The band structure of the Bi-terminated film is shown in Fig.~\ref{fig:hua1}b. The two helical states still robustly persist and shift below $E_F$, agreeing with experiments. Additionally, the degeneracy of the two helical states is lifted, of which the upper and lower bands correspond to the helical state at the upper (Bi-terminated) and lower (Se-terminated) surface, respectively.

We choose 2 monolayers (ML) of Au deposited on the Bi surface of the Bi$_2$Se$_3$ film (Fig.~\ref{fig:hua1}c inset) as a model system because of its optimal stability for subsequent calculations. Here 1 ML is defined to be the same number of atoms as that in each atomic layer of Bi$_2$Se$_3$, which is equal to 0.48 times the atom density in a (111) layer of bulk Au. Figs.~\ref{fig:hua1}c and d show the band structures of the Au-covered Bi$_2$Se$_3$ film without and with SOC, respectively. Two helical states  emerge only when the SOC is switched on, confirming that the helical state indeed originates from the SOC of the bulk states. This observation allows us to conveniently isolate the effects of the helical state by comparative studies with and without SOC. The shape of the two helical state bands near the $\Gamma$ point closely resembles that of the helical state in Fig.~\ref{fig:hua1}b for the Bi-terminated Bi$_2$Se$_3$ film despite the slight shift in their relative positions in energy. Therefore the helical state survives even if the Bi$_2$Se$_3$ surface is completely buried under the 2 ML Au film.

\begin{figure}[h]
	\includegraphics[width = 3.4 in]{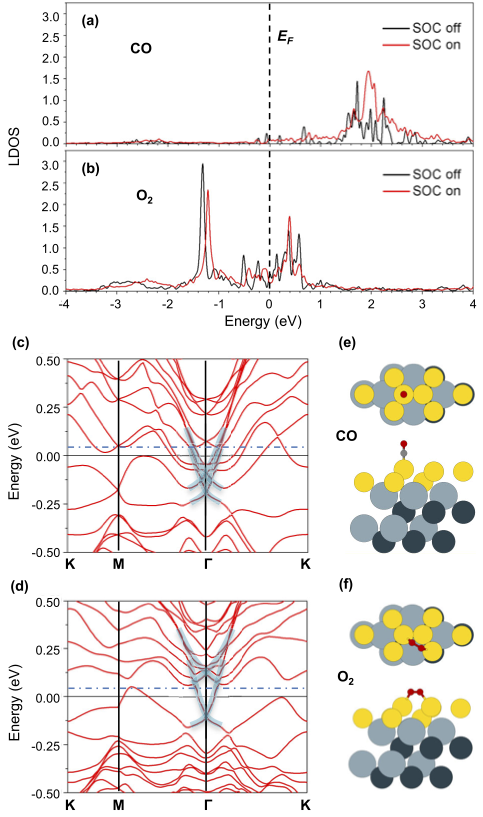}
	\caption{(a) and (b) LDOS on the C atom of CO, and one O atom of O$_2$, featuring the energy range corresponding to the 2$\pi^*$ states of CO and O$_2$, respectively. (c) and (d) Band structures of the CO and O$_2$ adsorbed 2 ML Au-Bi$_2$Se$_3$ film, shown in a reduced Brillouin zone corresponding to the 2$\times$2 surface unit cell (Fig.~\ref{fig:hua1} (a) inset). The helical state bands are highlighted by the transparent blue lines. The blue dot-dash lines indicate the position of the upper Dirac point in Fig.~\ref{fig:hua1} (d). The upper and lower panels in (e) and (f) are the top and side views of the atomic structures. Small red balls-O; small gray balls-C; yellow balls-Au; light blue balls-Bi; dark blue balls-Se.}
	\label{fig:hua2}
\end{figure}

We now present a comparative study showing that the binding of both CO and O$_2$ on Au-covered Bi$_2$Se$_3$ are enhanced due to the presence of the helical states. With SOC, the binding energy of CO is considerably enhanced by 0.2 eV compared to that without SOC, accompanied by a decrease of the C-Au bond length from 2.029 \r{A} to 1.981 \r{A}. The enhanced CO binding with SOC is due to the static electron transfer facilitated by the helical state. To see this effect we first compare the local density of states (LDOS) on the C atom of an adsorbed CO with and without SOC, shown in Fig.~\ref{fig:hua2}a. The antibonding 2$\pi^*$ states shift to higher energies with SOC, indicating decreased electron occupation, and hence enhanced CO-Au binding. On the other hand, from Fig.~\ref{fig:hua2}c, the top Dirac band, corresponding to the helical state on the Au-deposited Bi-terminated surface, shifts to lower energy after the adsorption of CO, indicating increased electron occupation. Taken together, net electrons are transferred to the helical state serving as an electron bath when CO is adsorbed on the surface. 

Next we show that the helical state as an electron bath can also enhance the adsorption of O$_2$, but by invoking a different direction of static electron transfer. On the Au-Bi$_2$Se$_3$ substrate, O$_2$ binding energy increases by 0.16 eV with SOC. The LDOS on one O atom of O$_2$ is shown in Fig.~\ref{fig:hua2}b. The two groups of peaks below and above $E_F$ correspond to the spin-up and spin-down antibonding 2$\pi^*$ states, respectively. As the half-filled 2$\pi^*$ states hybridize with the Au $d$ states, more electrons will be transferred to the 2$\pi^*$ states and promote O$_2$ toward dissociation. At the same time, the spin splitting of the 2$\pi^*$ states will decrease due to the weakened O-O bond. In Fig.~\ref{fig:hua2}b, both groups of the spin-splitted peaks shift toward $E_F$ after turning on SOC, indicating decreased spin splitting in the O$_2$ orbitals. Meanwhile, the O-O bond length increases from 1.289 \r{A} without SOC to 1.299 \r{A} with SOC. The increased electron occupation of the 2$\pi^*$ states upon switching on SOC is not easily visible from Fig.~\ref{fig:hua2}b, but is confirmed by the calculated increase in the relative spectral weight of the 2$\pi^*$ DOS below $E_F$, equal to 0.56 with SOC and 0.55 without SOC. This difference is roughly equal to 0.04 $e$ per O$_2$ molecule, originated from the helical state. On the other hand, from Fig.~\ref{fig:hua2}d, the top helical state Dirac band shifts upward compared to that without O$_2$ adsorption, indicating that electrons are transferred out of the helical state. Therefore, rather than accepting electrons as in the case of CO, the helical state now donates electrons and promotes O$_2$ toward dissociative adsorption on Au. Moreover, the adsorption energy of O$_2$ is now comparable to that of CO with a moderate strength, which is a desirable feature for easier reaction and high catalytic activity.

\section{Concluding remarks}
In the past several years, the TI-based heterostructures have attracted considerable attention due to many exotic properties related to the new discovered topological helical states. To achieve the long term goal to really utilize these properties in next-generation quantum devices, we should take an important step to control the spatial location of the helical states in the design structure. In this paper, we have introduced our recent efforts to explore the key parameters of the topological proximity effect, which is manifested by the shifting of the helical states in TI-based heterostructures. A rich phase diagram, depending on the SOC, band gap, and interface coupling, has been presented. We hope these encouraging results could contribute to stimulate more efforts along this direction. 


\begin{thebibliography}{50}
\expandafter\ifx\csname natexlab\endcsname\relax\def\natexlab#1{#1}\fi
\expandafter\ifx\csname bibnamefont\endcsname\relax
  \def\bibnamefont#1{#1}\fi
\expandafter\ifx\csname bibfnamefont\endcsname\relax
  \def\bibfnamefont#1{#1}\fi
\expandafter\ifx\csname citenamefont\endcsname\relax
  \def\citenamefont#1{#1}\fi
\expandafter\ifx\csname url\endcsname\relax
  \def\url#1{\texttt{#1}}\fi
\expandafter\ifx\csname urlprefix\endcsname\relax\def\urlprefix{URL }\fi
\providecommand{\bibinfo}[2]{#2}
\providecommand{\eprint}[2][]{\url{#2}}

\bibitem[{\citenamefont{Thouless et~al.}(1982)\citenamefont{Thouless, Kohmoto,
  Nightingale, and den Nijs}}]{PhysRevLett.49.405}
\bibinfo{author}{\bibfnamefont{D.~J.} \bibnamefont{Thouless}},
  \bibinfo{author}{\bibfnamefont{M.}~\bibnamefont{Kohmoto}},
  \bibinfo{author}{\bibfnamefont{M.~P.} \bibnamefont{Nightingale}},
  \bibnamefont{and} \bibinfo{author}{\bibfnamefont{M.}~\bibnamefont{den Nijs}},
  \bibinfo{journal}{Phys. Rev. Lett.} \textbf{\bibinfo{volume}{49}},
  \bibinfo{pages}{405} (\bibinfo{year}{1982}).

\bibitem[{\citenamefont{Hasan and Kane}(2010)}]{RevModPhys.82.3045}
\bibinfo{author}{\bibfnamefont{M.~Z.} \bibnamefont{Hasan}} \bibnamefont{and}
  \bibinfo{author}{\bibfnamefont{C.~L.} \bibnamefont{Kane}},
  \bibinfo{journal}{Rev. Mod. Phys.} \textbf{\bibinfo{volume}{82}},
  \bibinfo{pages}{3045} (\bibinfo{year}{2010}).

\bibitem[{\citenamefont{Qi and Zhang}(2011)}]{RevModPhys.83.1057}
\bibinfo{author}{\bibfnamefont{X.-L.} \bibnamefont{Qi}} \bibnamefont{and}
  \bibinfo{author}{\bibfnamefont{S.-C.} \bibnamefont{Zhang}},
  \bibinfo{journal}{Rev. Mod. Phys.} \textbf{\bibinfo{volume}{83}},
  \bibinfo{pages}{1057} (\bibinfo{year}{2011}).

\bibitem[{\citenamefont{Kane and
  Mele}(2005{\natexlab{a}})}]{PhysRevLett.95.146802}
\bibinfo{author}{\bibfnamefont{C.~L.} \bibnamefont{Kane}} \bibnamefont{and}
  \bibinfo{author}{\bibfnamefont{E.~J.} \bibnamefont{Mele}},
  \bibinfo{journal}{Phys. Rev. Lett.} \textbf{\bibinfo{volume}{95}},
  \bibinfo{pages}{146802} (\bibinfo{year}{2005}{\natexlab{a}}).

\bibitem[{\citenamefont{Kane and
  Mele}(2005{\natexlab{b}})}]{PhysRevLett.95.226801}
\bibinfo{author}{\bibfnamefont{C.~L.} \bibnamefont{Kane}} \bibnamefont{and}
  \bibinfo{author}{\bibfnamefont{E.~J.} \bibnamefont{Mele}},
  \bibinfo{journal}{Phys. Rev. Lett.} \textbf{\bibinfo{volume}{95}},
  \bibinfo{pages}{226801} (\bibinfo{year}{2005}{\natexlab{b}}).

\bibitem[{\citenamefont{Bernevig et~al.}(2006)\citenamefont{Bernevig, Hughes,
  and Zhang}}]{Bernevig:2006kx}
\bibinfo{author}{\bibfnamefont{B.~A.} \bibnamefont{Bernevig}},
  \bibinfo{author}{\bibfnamefont{T.~L.} \bibnamefont{Hughes}},
  \bibnamefont{and} \bibinfo{author}{\bibfnamefont{S.-C.} \bibnamefont{Zhang}},
  \bibinfo{journal}{Science} \textbf{\bibinfo{volume}{314}},
  \bibinfo{pages}{1757} (\bibinfo{year}{2006}).

\bibitem[{\citenamefont{K{\"o}nig et~al.}(2007)\citenamefont{K{\"o}nig,
  Wiedmann, Br{\"u}ne, Roth, Buhmann, Molenkamp, Qi, and Zhang}}]{Konig:2007uq}
\bibinfo{author}{\bibfnamefont{M.}~\bibnamefont{K{\"o}nig}},
  \bibinfo{author}{\bibfnamefont{S.}~\bibnamefont{Wiedmann}},
  \bibinfo{author}{\bibfnamefont{C.}~\bibnamefont{Br{\"u}ne}},
  \bibinfo{author}{\bibfnamefont{A.}~\bibnamefont{Roth}},
  \bibinfo{author}{\bibfnamefont{H.}~\bibnamefont{Buhmann}},
  \bibinfo{author}{\bibfnamefont{L.~W.} \bibnamefont{Molenkamp}},
  \bibinfo{author}{\bibfnamefont{X.-L.} \bibnamefont{Qi}}, \bibnamefont{and}
  \bibinfo{author}{\bibfnamefont{S.-C.} \bibnamefont{Zhang}},
  \bibinfo{journal}{Science} \textbf{\bibinfo{volume}{318}},
  \bibinfo{pages}{766} (\bibinfo{year}{2007}).

\bibitem[{\citenamefont{Kong and Cui}(2011)}]{Kong:2011tg}
\bibinfo{author}{\bibfnamefont{D.}~\bibnamefont{Kong}} \bibnamefont{and}
  \bibinfo{author}{\bibfnamefont{Y.}~\bibnamefont{Cui}}, \bibinfo{journal}{Nat
  Chem} \textbf{\bibinfo{volume}{3}}, \bibinfo{pages}{845}
  (\bibinfo{year}{2011}).

\bibitem[{\citenamefont{Fu and Kane}(2008)}]{PhysRevLett.100.096407}
\bibinfo{author}{\bibfnamefont{L.}~\bibnamefont{Fu}} \bibnamefont{and}
  \bibinfo{author}{\bibfnamefont{C.~L.} \bibnamefont{Kane}},
  \bibinfo{journal}{Phys. Rev. Lett.} \textbf{\bibinfo{volume}{100}},
  \bibinfo{pages}{096407} (\bibinfo{year}{2008}).

\bibitem[{\citenamefont{Stanescu et~al.}(2010)\citenamefont{Stanescu, Sau,
  Lutchyn, and Das~Sarma}}]{PhysRevB.81.241310}
\bibinfo{author}{\bibfnamefont{T.~D.} \bibnamefont{Stanescu}},
  \bibinfo{author}{\bibfnamefont{J.~D.} \bibnamefont{Sau}},
  \bibinfo{author}{\bibfnamefont{R.~M.} \bibnamefont{Lutchyn}},
  \bibnamefont{and}
  \bibinfo{author}{\bibfnamefont{S.}~\bibnamefont{Das~Sarma}},
  \bibinfo{journal}{Phys. Rev. B} \textbf{\bibinfo{volume}{81}},
  \bibinfo{pages}{241310} (\bibinfo{year}{2010}).

\bibitem[{\citenamefont{Mahfouzi et~al.}(2010)\citenamefont{Mahfouzi,
  Nikoli\ifmmode~\acute{c}\else \'{c}\fi{}, Chen, and
  Chang}}]{PhysRevB.82.195440}
\bibinfo{author}{\bibfnamefont{F.}~\bibnamefont{Mahfouzi}},
  \bibinfo{author}{\bibfnamefont{B.~K.}
  \bibnamefont{Nikoli\ifmmode~\acute{c}\else \'{c}\fi{}}},
  \bibinfo{author}{\bibfnamefont{S.-H.} \bibnamefont{Chen}}, \bibnamefont{and}
  \bibinfo{author}{\bibfnamefont{C.-R.} \bibnamefont{Chang}},
  \bibinfo{journal}{Phys. Rev. B} \textbf{\bibinfo{volume}{82}},
  \bibinfo{pages}{195440} (\bibinfo{year}{2010}).

\bibitem[{\citenamefont{Garate and Franz}(2010)}]{PhysRevLett.104.146802}
\bibinfo{author}{\bibfnamefont{I.}~\bibnamefont{Garate}} \bibnamefont{and}
  \bibinfo{author}{\bibfnamefont{M.}~\bibnamefont{Franz}},
  \bibinfo{journal}{Phys. Rev. Lett.} \textbf{\bibinfo{volume}{104}},
  \bibinfo{pages}{146802} (\bibinfo{year}{2010}).

\bibitem[{\citenamefont{Chang et~al.}(2011)\citenamefont{Chang, Jadaun,
  Register, Banerjee, and Sahu}}]{PhysRevB.84.155105}
\bibinfo{author}{\bibfnamefont{J.}~\bibnamefont{Chang}},
  \bibinfo{author}{\bibfnamefont{P.}~\bibnamefont{Jadaun}},
  \bibinfo{author}{\bibfnamefont{L.~F.} \bibnamefont{Register}},
  \bibinfo{author}{\bibfnamefont{S.~K.} \bibnamefont{Banerjee}},
  \bibnamefont{and} \bibinfo{author}{\bibfnamefont{B.}~\bibnamefont{Sahu}},
  \bibinfo{journal}{Phys. Rev. B} \textbf{\bibinfo{volume}{84}},
  \bibinfo{pages}{155105} (\bibinfo{year}{2011}).

\bibitem[{\citenamefont{Hutasoit and Stanescu}(2011)}]{PhysRevB.84.085103}
\bibinfo{author}{\bibfnamefont{J.~A.} \bibnamefont{Hutasoit}} \bibnamefont{and}
  \bibinfo{author}{\bibfnamefont{T.~D.} \bibnamefont{Stanescu}},
  \bibinfo{journal}{Phys. Rev. B} \textbf{\bibinfo{volume}{84}},
  \bibinfo{pages}{085103} (\bibinfo{year}{2011}).

\bibitem[{\citenamefont{Hirahara et~al.}(2011)\citenamefont{Hirahara,
  Bihlmayer, Sakamoto, Yamada, Miyazaki, Kimura, Bl\"ugel, and
  Hasegawa}}]{PhysRevLett.107.166801}
\bibinfo{author}{\bibfnamefont{T.}~\bibnamefont{Hirahara}},
  \bibinfo{author}{\bibfnamefont{G.}~\bibnamefont{Bihlmayer}},
  \bibinfo{author}{\bibfnamefont{Y.}~\bibnamefont{Sakamoto}},
  \bibinfo{author}{\bibfnamefont{M.}~\bibnamefont{Yamada}},
  \bibinfo{author}{\bibfnamefont{H.}~\bibnamefont{Miyazaki}},
  \bibinfo{author}{\bibfnamefont{S.-i.} \bibnamefont{Kimura}},
  \bibinfo{author}{\bibfnamefont{S.}~\bibnamefont{Bl\"ugel}}, \bibnamefont{and}
  \bibinfo{author}{\bibfnamefont{S.}~\bibnamefont{Hasegawa}},
  \bibinfo{journal}{Phys. Rev. Lett.} \textbf{\bibinfo{volume}{107}},
  \bibinfo{pages}{166801} (\bibinfo{year}{2011}).

\bibitem[{\citenamefont{Chen et~al.}(2011)\citenamefont{Chen, Zhu, Xiao, and
  Zhang}}]{PhysRevLett.107.056804}
\bibinfo{author}{\bibfnamefont{H.}~\bibnamefont{Chen}},
  \bibinfo{author}{\bibfnamefont{W.}~\bibnamefont{Zhu}},
  \bibinfo{author}{\bibfnamefont{D.}~\bibnamefont{Xiao}}, \bibnamefont{and}
  \bibinfo{author}{\bibfnamefont{Z.}~\bibnamefont{Zhang}},
  \bibinfo{journal}{Phys. Rev. Lett.} \textbf{\bibinfo{volume}{107}},
  \bibinfo{pages}{056804} (\bibinfo{year}{2011}).

\bibitem[{\citenamefont{Cook and Franz}(2011)}]{PhysRevB.84.201105}
\bibinfo{author}{\bibfnamefont{A.}~\bibnamefont{Cook}} \bibnamefont{and}
  \bibinfo{author}{\bibfnamefont{M.}~\bibnamefont{Franz}},
  \bibinfo{journal}{Phys. Rev. B} \textbf{\bibinfo{volume}{84}},
  \bibinfo{pages}{201105} (\bibinfo{year}{2011}).

\bibitem[{\citenamefont{Zhang et~al.}(2012)\citenamefont{Zhang, Zhang, Zhu,
  Schwingenschl{\"o}gl, and Cui}}]{Zhang:2012fk}
\bibinfo{author}{\bibfnamefont{Q.}~\bibnamefont{Zhang}},
  \bibinfo{author}{\bibfnamefont{Z.}~\bibnamefont{Zhang}},
  \bibinfo{author}{\bibfnamefont{Z.}~\bibnamefont{Zhu}},
  \bibinfo{author}{\bibfnamefont{U.}~\bibnamefont{Schwingenschl{\"o}gl}},
  \bibnamefont{and} \bibinfo{author}{\bibfnamefont{Y.}~\bibnamefont{Cui}},
  \bibinfo{journal}{ACS Nano} \textbf{\bibinfo{volume}{6}},
  \bibinfo{pages}{2345} (\bibinfo{year}{2012}).

\bibitem[{\citenamefont{Nakayama et~al.}(2012)\citenamefont{Nakayama, Eto,
  Tanaka, Sato, Souma, Takahashi, Segawa, and Ando}}]{PhysRevLett.109.236804}
\bibinfo{author}{\bibfnamefont{K.}~\bibnamefont{Nakayama}},
  \bibinfo{author}{\bibfnamefont{K.}~\bibnamefont{Eto}},
  \bibinfo{author}{\bibfnamefont{Y.}~\bibnamefont{Tanaka}},
  \bibinfo{author}{\bibfnamefont{T.}~\bibnamefont{Sato}},
  \bibinfo{author}{\bibfnamefont{S.}~\bibnamefont{Souma}},
  \bibinfo{author}{\bibfnamefont{T.}~\bibnamefont{Takahashi}},
  \bibinfo{author}{\bibfnamefont{K.}~\bibnamefont{Segawa}}, \bibnamefont{and}
  \bibinfo{author}{\bibfnamefont{Y.}~\bibnamefont{Ando}},
  \bibinfo{journal}{Phys. Rev. Lett.} \textbf{\bibinfo{volume}{109}},
  \bibinfo{pages}{236804} (\bibinfo{year}{2012}).

\bibitem[{\citenamefont{Shevtsov et~al.}(2012)\citenamefont{Shevtsov, Carmier,
  Petitjean, Groth, Carpentier, and Waintal}}]{PhysRevX.2.031004}
\bibinfo{author}{\bibfnamefont{O.}~\bibnamefont{Shevtsov}},
  \bibinfo{author}{\bibfnamefont{P.}~\bibnamefont{Carmier}},
  \bibinfo{author}{\bibfnamefont{C.}~\bibnamefont{Petitjean}},
  \bibinfo{author}{\bibfnamefont{C.}~\bibnamefont{Groth}},
  \bibinfo{author}{\bibfnamefont{D.}~\bibnamefont{Carpentier}},
  \bibnamefont{and} \bibinfo{author}{\bibfnamefont{X.}~\bibnamefont{Waintal}},
  \bibinfo{journal}{Phys. Rev. X} \textbf{\bibinfo{volume}{2}},
  \bibinfo{pages}{031004} (\bibinfo{year}{2012}).

\bibitem[{\citenamefont{Culcer}(2012)}]{Culcer:2012oq}
\bibinfo{author}{\bibfnamefont{D.}~\bibnamefont{Culcer}},
  \bibinfo{journal}{Physica E: Low-dimensional Systems and Nanostructures}
  \textbf{\bibinfo{volume}{44}}, \bibinfo{pages}{860} (\bibinfo{year}{2012}).

\bibitem[{\citenamefont{Qu et~al.}(2012)\citenamefont{Qu, Yang, Shen, Ding,
  Chen, Ji, Liu, Fan, Jing, Yang et~al.}}]{Qu:2012qf}
\bibinfo{author}{\bibfnamefont{F.}~\bibnamefont{Qu}},
  \bibinfo{author}{\bibfnamefont{F.}~\bibnamefont{Yang}},
  \bibinfo{author}{\bibfnamefont{J.}~\bibnamefont{Shen}},
  \bibinfo{author}{\bibfnamefont{Y.}~\bibnamefont{Ding}},
  \bibinfo{author}{\bibfnamefont{J.}~\bibnamefont{Chen}},
  \bibinfo{author}{\bibfnamefont{Z.}~\bibnamefont{Ji}},
  \bibinfo{author}{\bibfnamefont{G.}~\bibnamefont{Liu}},
  \bibinfo{author}{\bibfnamefont{J.}~\bibnamefont{Fan}},
  \bibinfo{author}{\bibfnamefont{X.}~\bibnamefont{Jing}},
  \bibinfo{author}{\bibfnamefont{C.}~\bibnamefont{Yang}}, \bibnamefont{et~al.},
  \bibinfo{journal}{Sci. Rep.} \textbf{\bibinfo{volume}{2}}
  (\bibinfo{year}{2012}).

\bibitem[{\citenamefont{{Zhang} et~al.}(2012)\citenamefont{{Zhang}, {Li}, {Wu},
  {Wang}, {Culcer}, {Kaxiras}, and {Zhang}}}]{2012arXiv1212.1343Z}
\bibinfo{author}{\bibfnamefont{G.}~\bibnamefont{{Zhang}}},
  \bibinfo{author}{\bibfnamefont{X.}~\bibnamefont{{Li}}},
  \bibinfo{author}{\bibfnamefont{G.}~\bibnamefont{{Wu}}},
  \bibinfo{author}{\bibfnamefont{J.}~\bibnamefont{{Wang}}},
  \bibinfo{author}{\bibfnamefont{D.}~\bibnamefont{{Culcer}}},
  \bibinfo{author}{\bibfnamefont{E.}~\bibnamefont{{Kaxiras}}},
  \bibnamefont{and} \bibinfo{author}{\bibfnamefont{Z.}~\bibnamefont{{Zhang}}},
  \bibinfo{journal}{ArXiv e-prints}  (\bibinfo{year}{2012}),
  \eprint{1212.1343}.

\bibitem[{\citenamefont{{Eremeev} et~al.}(2013)\citenamefont{{Eremeev},
  {Men'shov}, {Tugushev}, {Echenique}, and {Chulkov}}}]{2013arXiv1304.1275E}
\bibinfo{author}{\bibfnamefont{S.~V.} \bibnamefont{{Eremeev}}},
  \bibinfo{author}{\bibfnamefont{V.~N.} \bibnamefont{{Men'shov}}},
  \bibinfo{author}{\bibfnamefont{V.~V.} \bibnamefont{{Tugushev}}},
  \bibinfo{author}{\bibfnamefont{P.~M.} \bibnamefont{{Echenique}}},
  \bibnamefont{and} \bibinfo{author}{\bibfnamefont{E.~V.}
  \bibnamefont{{Chulkov}}}, \bibinfo{journal}{ArXiv e-prints}
  (\bibinfo{year}{2013}), \eprint{1304.1275}.

\bibitem[{\citenamefont{Ueda et~al.}(2013)\citenamefont{Ueda, Kawakami, and
  Sigrist}}]{PhysRevB.87.161108}
\bibinfo{author}{\bibfnamefont{S.}~\bibnamefont{Ueda}},
  \bibinfo{author}{\bibfnamefont{N.}~\bibnamefont{Kawakami}}, \bibnamefont{and}
  \bibinfo{author}{\bibfnamefont{M.}~\bibnamefont{Sigrist}},
  \bibinfo{journal}{Phys. Rev. B} \textbf{\bibinfo{volume}{87}},
  \bibinfo{pages}{161108} (\bibinfo{year}{2013}).

\bibitem[{\citenamefont{Luo and Qi}(2013)}]{PhysRevB.87.085431}
\bibinfo{author}{\bibfnamefont{W.}~\bibnamefont{Luo}} \bibnamefont{and}
  \bibinfo{author}{\bibfnamefont{X.-L.} \bibnamefont{Qi}},
  \bibinfo{journal}{Phys. Rev. B} \textbf{\bibinfo{volume}{87}},
  \bibinfo{pages}{085431} (\bibinfo{year}{2013}).

\bibitem[{\citenamefont{Wu et~al.}(2013)\citenamefont{Wu, Chen, Sun, Li, Cui,
  Franchini, Wang, Chen, and Zhang}}]{Wu:2013fk}
\bibinfo{author}{\bibfnamefont{G.}~\bibnamefont{Wu}},
  \bibinfo{author}{\bibfnamefont{H.}~\bibnamefont{Chen}},
  \bibinfo{author}{\bibfnamefont{Y.}~\bibnamefont{Sun}},
  \bibinfo{author}{\bibfnamefont{X.}~\bibnamefont{Li}},
  \bibinfo{author}{\bibfnamefont{P.}~\bibnamefont{Cui}},
  \bibinfo{author}{\bibfnamefont{C.}~\bibnamefont{Franchini}},
  \bibinfo{author}{\bibfnamefont{J.}~\bibnamefont{Wang}},
  \bibinfo{author}{\bibfnamefont{X.-Q.} \bibnamefont{Chen}}, \bibnamefont{and}
  \bibinfo{author}{\bibfnamefont{Z.}~\bibnamefont{Zhang}},
  \bibinfo{journal}{Sci. Rep.} \textbf{\bibinfo{volume}{3}}
  (\bibinfo{year}{2013}).

\bibitem[{\citenamefont{Bj\"ornson and
  Black-Schaffer}(2013)}]{PhysRevB.88.024501}
\bibinfo{author}{\bibfnamefont{K.}~\bibnamefont{Bj\"ornson}} \bibnamefont{and}
  \bibinfo{author}{\bibfnamefont{A.~M.} \bibnamefont{Black-Schaffer}},
  \bibinfo{journal}{Phys. Rev. B} \textbf{\bibinfo{volume}{88}},
  \bibinfo{pages}{024501} (\bibinfo{year}{2013}).

\bibitem[{\citenamefont{Nayak et~al.}(2008)\citenamefont{Nayak, Simon, Stern,
  Freedman, and Das~Sarma}}]{RevModPhys.80.1083}
\bibinfo{author}{\bibfnamefont{C.}~\bibnamefont{Nayak}},
  \bibinfo{author}{\bibfnamefont{S.~H.} \bibnamefont{Simon}},
  \bibinfo{author}{\bibfnamefont{A.}~\bibnamefont{Stern}},
  \bibinfo{author}{\bibfnamefont{M.}~\bibnamefont{Freedman}}, \bibnamefont{and}
  \bibinfo{author}{\bibfnamefont{S.}~\bibnamefont{Das~Sarma}},
  \bibinfo{journal}{Rev. Mod. Phys.} \textbf{\bibinfo{volume}{80}},
  \bibinfo{pages}{1083} (\bibinfo{year}{2008}).

\bibitem[{\citenamefont{Qi et~al.}(2009)\citenamefont{Qi, Li, Zang, and
  Zhang}}]{Qi:2009nx}
\bibinfo{author}{\bibfnamefont{X.-L.} \bibnamefont{Qi}},
  \bibinfo{author}{\bibfnamefont{R.}~\bibnamefont{Li}},
  \bibinfo{author}{\bibfnamefont{J.}~\bibnamefont{Zang}}, \bibnamefont{and}
  \bibinfo{author}{\bibfnamefont{S.-C.} \bibnamefont{Zhang}},
  \bibinfo{journal}{Science} \textbf{\bibinfo{volume}{323}},
  \bibinfo{pages}{1184} (\bibinfo{year}{2009}).

\bibitem[{\citenamefont{Zhang et~al.}(2010{\natexlab{a}})\citenamefont{Zhang,
  Yu, Zhang, Dai, and Fang}}]{1367-2630-12-6-065013}
\bibinfo{author}{\bibfnamefont{W.}~\bibnamefont{Zhang}},
  \bibinfo{author}{\bibfnamefont{R.}~\bibnamefont{Yu}},
  \bibinfo{author}{\bibfnamefont{H.-J.} \bibnamefont{Zhang}},
  \bibinfo{author}{\bibfnamefont{X.}~\bibnamefont{Dai}}, \bibnamefont{and}
  \bibinfo{author}{\bibfnamefont{Z.}~\bibnamefont{Fang}}, \bibinfo{journal}{New
  Journal of Physics} \textbf{\bibinfo{volume}{12}}, \bibinfo{pages}{065013}
  (\bibinfo{year}{2010}{\natexlab{a}}).

\bibitem[{\citenamefont{Eremeev et~al.}(2012)\citenamefont{Eremeev, Landolt,
  Menshchikova, Slomski, Koroteev, Aliev, Babanly, Henk, Ernst, Patthey
  et~al.}}]{Eremeev:2012xi}
\bibinfo{author}{\bibfnamefont{S.~V.} \bibnamefont{Eremeev}},
  \bibinfo{author}{\bibfnamefont{G.}~\bibnamefont{Landolt}},
  \bibinfo{author}{\bibfnamefont{T.~V.} \bibnamefont{Menshchikova}},
  \bibinfo{author}{\bibfnamefont{B.}~\bibnamefont{Slomski}},
  \bibinfo{author}{\bibfnamefont{Y.~M.} \bibnamefont{Koroteev}},
  \bibinfo{author}{\bibfnamefont{Z.~S.} \bibnamefont{Aliev}},
  \bibinfo{author}{\bibfnamefont{M.~B.} \bibnamefont{Babanly}},
  \bibinfo{author}{\bibfnamefont{J.}~\bibnamefont{Henk}},
  \bibinfo{author}{\bibfnamefont{A.}~\bibnamefont{Ernst}},
  \bibinfo{author}{\bibfnamefont{L.}~\bibnamefont{Patthey}},
  \bibnamefont{et~al.}, \bibinfo{journal}{Nat Commun}
  \textbf{\bibinfo{volume}{3}}, \bibinfo{pages}{635} (\bibinfo{year}{2012}).

\bibitem[{\citenamefont{Black-Schaffer and
  Balatsky}(2012)}]{PhysRevB.85.121103}
\bibinfo{author}{\bibfnamefont{A.~M.} \bibnamefont{Black-Schaffer}}
  \bibnamefont{and} \bibinfo{author}{\bibfnamefont{A.~V.}
  \bibnamefont{Balatsky}}, \bibinfo{journal}{Phys. Rev. B}
  \textbf{\bibinfo{volume}{85}}, \bibinfo{pages}{121103}
  (\bibinfo{year}{2012}).

\bibitem[{\citenamefont{Xu et~al.}(2011)\citenamefont{Xu, Xia, Wray, Jia,
  Meier, Dil, Osterwalder, Slomski, Bansil, Lin et~al.}}]{Xu:2011ve}
\bibinfo{author}{\bibfnamefont{S.-Y.} \bibnamefont{Xu}},
  \bibinfo{author}{\bibfnamefont{Y.}~\bibnamefont{Xia}},
  \bibinfo{author}{\bibfnamefont{L.~A.} \bibnamefont{Wray}},
  \bibinfo{author}{\bibfnamefont{S.}~\bibnamefont{Jia}},
  \bibinfo{author}{\bibfnamefont{F.}~\bibnamefont{Meier}},
  \bibinfo{author}{\bibfnamefont{J.~H.} \bibnamefont{Dil}},
  \bibinfo{author}{\bibfnamefont{J.}~\bibnamefont{Osterwalder}},
  \bibinfo{author}{\bibfnamefont{B.}~\bibnamefont{Slomski}},
  \bibinfo{author}{\bibfnamefont{A.}~\bibnamefont{Bansil}},
  \bibinfo{author}{\bibfnamefont{H.}~\bibnamefont{Lin}}, \bibnamefont{et~al.},
  \bibinfo{journal}{Science} \textbf{\bibinfo{volume}{332}},
  \bibinfo{pages}{560} (\bibinfo{year}{2011}).

\bibitem[{\citenamefont{Weeks et~al.}(2011)\citenamefont{Weeks, Hu, Alicea,
  Franz, and Wu}}]{PhysRevX.1.021001}
\bibinfo{author}{\bibfnamefont{C.}~\bibnamefont{Weeks}},
  \bibinfo{author}{\bibfnamefont{J.}~\bibnamefont{Hu}},
  \bibinfo{author}{\bibfnamefont{J.}~\bibnamefont{Alicea}},
  \bibinfo{author}{\bibfnamefont{M.}~\bibnamefont{Franz}}, \bibnamefont{and}
  \bibinfo{author}{\bibfnamefont{R.}~\bibnamefont{Wu}}, \bibinfo{journal}{Phys.
  Rev. X} \textbf{\bibinfo{volume}{1}}, \bibinfo{pages}{021001}
  (\bibinfo{year}{2011}).

\bibitem[{\citenamefont{Chadov et~al.}(2010)\citenamefont{Chadov, Qi,
  K{\~A}¼bler, Fecher, Felser, and Zhang}}]{Chadov:2010mq}
\bibinfo{author}{\bibfnamefont{S.}~\bibnamefont{Chadov}},
  \bibinfo{author}{\bibfnamefont{X.}~\bibnamefont{Qi}},
  \bibinfo{author}{\bibfnamefont{J.}~\bibnamefont{K{\~A}¼bler}},
  \bibinfo{author}{\bibfnamefont{G.~H.} \bibnamefont{Fecher}},
  \bibinfo{author}{\bibfnamefont{C.}~\bibnamefont{Felser}}, \bibnamefont{and}
  \bibinfo{author}{\bibfnamefont{S.~C.} \bibnamefont{Zhang}},
  \bibinfo{journal}{Nature Mater.} \textbf{\bibinfo{volume}{9}},
  \bibinfo{pages}{541} (\bibinfo{year}{2010}).

\bibitem[{\citenamefont{Xiao et~al.}(2010)\citenamefont{Xiao, Yao, Feng, Wen,
  Zhu, Chen, Stocks, and Zhang}}]{PhysRevLett.105.096404}
\bibinfo{author}{\bibfnamefont{D.}~\bibnamefont{Xiao}},
  \bibinfo{author}{\bibfnamefont{Y.}~\bibnamefont{Yao}},
  \bibinfo{author}{\bibfnamefont{W.}~\bibnamefont{Feng}},
  \bibinfo{author}{\bibfnamefont{J.}~\bibnamefont{Wen}},
  \bibinfo{author}{\bibfnamefont{W.}~\bibnamefont{Zhu}},
  \bibinfo{author}{\bibfnamefont{X.-Q.} \bibnamefont{Chen}},
  \bibinfo{author}{\bibfnamefont{G.~M.} \bibnamefont{Stocks}},
  \bibnamefont{and} \bibinfo{author}{\bibfnamefont{Z.}~\bibnamefont{Zhang}},
  \bibinfo{journal}{Phys. Rev. Lett.} \textbf{\bibinfo{volume}{105}},
  \bibinfo{pages}{096404} (\bibinfo{year}{2010}).

\bibitem[{\citenamefont{Lin et~al.}(2010)\citenamefont{Lin, Wray, Xia, Xu, Jia,
  Cava, Bansil, and Hasan}}]{Lin:2010wm}
\bibinfo{author}{\bibfnamefont{H.}~\bibnamefont{Lin}},
  \bibinfo{author}{\bibfnamefont{L.~A.} \bibnamefont{Wray}},
  \bibinfo{author}{\bibfnamefont{Y.}~\bibnamefont{Xia}},
  \bibinfo{author}{\bibfnamefont{S.}~\bibnamefont{Xu}},
  \bibinfo{author}{\bibfnamefont{S.}~\bibnamefont{Jia}},
  \bibinfo{author}{\bibfnamefont{R.~J.} \bibnamefont{Cava}},
  \bibinfo{author}{\bibfnamefont{A.}~\bibnamefont{Bansil}}, \bibnamefont{and}
  \bibinfo{author}{\bibfnamefont{M.~Z.} \bibnamefont{Hasan}},
  \bibinfo{journal}{Nature Mater.} \textbf{\bibinfo{volume}{9}},
  \bibinfo{pages}{546} (\bibinfo{year}{2010}).

\bibitem[{\citenamefont{Kim et~al.}(2012)\citenamefont{Kim, Kim, Kim, and
  Ihm}}]{Kim:2012ly}
\bibinfo{author}{\bibfnamefont{M.}~\bibnamefont{Kim}},
  \bibinfo{author}{\bibfnamefont{C.~H.} \bibnamefont{Kim}},
  \bibinfo{author}{\bibfnamefont{H.-S.} \bibnamefont{Kim}}, \bibnamefont{and}
  \bibinfo{author}{\bibfnamefont{J.}~\bibnamefont{Ihm}},
  \bibinfo{journal}{Proceedings of the National Academy of Sciences}
  \textbf{\bibinfo{volume}{109}}, \bibinfo{pages}{671} (\bibinfo{year}{2012}).

\bibitem[{\citenamefont{Qi and Zhang}(2010)}]{qi:33}
\bibinfo{author}{\bibfnamefont{X.-L.} \bibnamefont{Qi}} \bibnamefont{and}
  \bibinfo{author}{\bibfnamefont{S.-C.} \bibnamefont{Zhang}},
  \bibinfo{journal}{Physics Today} \textbf{\bibinfo{volume}{63}},
  \bibinfo{pages}{33} (\bibinfo{year}{2010}).

\bibitem[{\citenamefont{Lenz et~al.}(2007)\citenamefont{Lenz, Zander, and
  Kuch}}]{PhysRevLett.98.237201}
\bibinfo{author}{\bibfnamefont{K.}~\bibnamefont{Lenz}},
  \bibinfo{author}{\bibfnamefont{S.}~\bibnamefont{Zander}}, \bibnamefont{and}
  \bibinfo{author}{\bibfnamefont{W.}~\bibnamefont{Kuch}},
  \bibinfo{journal}{Phys. Rev. Lett.} \textbf{\bibinfo{volume}{98}},
  \bibinfo{pages}{237201} (\bibinfo{year}{2007}).

\bibitem[{\citenamefont{Helmes et~al.}(2008)\citenamefont{Helmes, Costi, and
  Rosch}}]{PhysRevLett.101.066802}
\bibinfo{author}{\bibfnamefont{R.~W.} \bibnamefont{Helmes}},
  \bibinfo{author}{\bibfnamefont{T.~A.} \bibnamefont{Costi}}, \bibnamefont{and}
  \bibinfo{author}{\bibfnamefont{A.}~\bibnamefont{Rosch}},
  \bibinfo{journal}{Phys. Rev. Lett.} \textbf{\bibinfo{volume}{101}},
  \bibinfo{pages}{066802} (\bibinfo{year}{2008}).

\bibitem[{\citenamefont{Yao et~al.}(2007)\citenamefont{Yao, Ye, Qi, Zhang, and
  Fang}}]{PhysRevB.75.041401}
\bibinfo{author}{\bibfnamefont{Y.}~\bibnamefont{Yao}},
  \bibinfo{author}{\bibfnamefont{F.}~\bibnamefont{Ye}},
  \bibinfo{author}{\bibfnamefont{X.-L.} \bibnamefont{Qi}},
  \bibinfo{author}{\bibfnamefont{S.-C.} \bibnamefont{Zhang}}, \bibnamefont{and}
  \bibinfo{author}{\bibfnamefont{Z.}~\bibnamefont{Fang}},
  \bibinfo{journal}{Phys. Rev. B} \textbf{\bibinfo{volume}{75}},
  \bibinfo{pages}{041401} (\bibinfo{year}{2007}).

\bibitem[{\citenamefont{Neto et~al.}(2009)\citenamefont{Neto, Guinea, Peres,
  Novoselov, and Geim}}]{neto:109}
\bibinfo{author}{\bibfnamefont{A.~H.~C.} \bibnamefont{Neto}},
  \bibinfo{author}{\bibfnamefont{F.}~\bibnamefont{Guinea}},
  \bibinfo{author}{\bibfnamefont{N.~M.~R.} \bibnamefont{Peres}},
  \bibinfo{author}{\bibfnamefont{K.~S.} \bibnamefont{Novoselov}},
  \bibnamefont{and} \bibinfo{author}{\bibfnamefont{A.~K.} \bibnamefont{Geim}},
  \bibinfo{journal}{Rev. Mod. Phys.} \textbf{\bibinfo{volume}{81}},
  \bibinfo{pages}{109} (\bibinfo{year}{2009}).

\bibitem[{\citenamefont{Fu et~al.}(2007)\citenamefont{Fu, Kane, and
  Mele}}]{PhysRevLett.98.106803}
\bibinfo{author}{\bibfnamefont{L.}~\bibnamefont{Fu}},
  \bibinfo{author}{\bibfnamefont{C.~L.} \bibnamefont{Kane}}, \bibnamefont{and}
  \bibinfo{author}{\bibfnamefont{E.~J.} \bibnamefont{Mele}},
  \bibinfo{journal}{Phys. Rev. Lett.} \textbf{\bibinfo{volume}{98}},
  \bibinfo{pages}{106803} (\bibinfo{year}{2007}).

\bibitem[{\citenamefont{Hsieh et~al.}(2008)\citenamefont{Hsieh, Qian, Wray,
  Xia, Hor, Cava, and Hasan}}]{Hsieh:2008vn}
\bibinfo{author}{\bibfnamefont{D.}~\bibnamefont{Hsieh}},
  \bibinfo{author}{\bibfnamefont{D.}~\bibnamefont{Qian}},
  \bibinfo{author}{\bibfnamefont{L.}~\bibnamefont{Wray}},
  \bibinfo{author}{\bibfnamefont{Y.}~\bibnamefont{Xia}},
  \bibinfo{author}{\bibfnamefont{Y.~S.} \bibnamefont{Hor}},
  \bibinfo{author}{\bibfnamefont{R.~J.} \bibnamefont{Cava}}, \bibnamefont{and}
  \bibinfo{author}{\bibfnamefont{M.~Z.} \bibnamefont{Hasan}},
  \bibinfo{journal}{Nature} \textbf{\bibinfo{volume}{452}},
  \bibinfo{pages}{970} (\bibinfo{year}{2008}).

\bibitem[{\citenamefont{Zhang et~al.}(2009)\citenamefont{Zhang, Cheng, Chen,
  Jia, Ma, He, Wang, Zhang, Dai, Fang et~al.}}]{PhysRevLett.103.266803}
\bibinfo{author}{\bibfnamefont{T.}~\bibnamefont{Zhang}},
  \bibinfo{author}{\bibfnamefont{P.}~\bibnamefont{Cheng}},
  \bibinfo{author}{\bibfnamefont{X.}~\bibnamefont{Chen}},
  \bibinfo{author}{\bibfnamefont{J.-F.} \bibnamefont{Jia}},
  \bibinfo{author}{\bibfnamefont{X.}~\bibnamefont{Ma}},
  \bibinfo{author}{\bibfnamefont{K.}~\bibnamefont{He}},
  \bibinfo{author}{\bibfnamefont{L.}~\bibnamefont{Wang}},
  \bibinfo{author}{\bibfnamefont{H.}~\bibnamefont{Zhang}},
  \bibinfo{author}{\bibfnamefont{X.}~\bibnamefont{Dai}},
  \bibinfo{author}{\bibfnamefont{Z.}~\bibnamefont{Fang}}, \bibnamefont{et~al.},
  \bibinfo{journal}{Phys. Rev. Lett.} \textbf{\bibinfo{volume}{103}},
  \bibinfo{pages}{266803} (\bibinfo{year}{2009}).

\bibitem[{\citenamefont{Hsieh et~al.}(2009)\citenamefont{Hsieh, Xia, Qian,
  Wray, Dil, Meier, Osterwalder, Patthey, Checkelsky, Ong
  et~al.}}]{Hsieh:2009lp}
\bibinfo{author}{\bibfnamefont{D.}~\bibnamefont{Hsieh}},
  \bibinfo{author}{\bibfnamefont{Y.}~\bibnamefont{Xia}},
  \bibinfo{author}{\bibfnamefont{D.}~\bibnamefont{Qian}},
  \bibinfo{author}{\bibfnamefont{L.}~\bibnamefont{Wray}},
  \bibinfo{author}{\bibfnamefont{J.~H.} \bibnamefont{Dil}},
  \bibinfo{author}{\bibfnamefont{F.}~\bibnamefont{Meier}},
  \bibinfo{author}{\bibfnamefont{J.}~\bibnamefont{Osterwalder}},
  \bibinfo{author}{\bibfnamefont{L.}~\bibnamefont{Patthey}},
  \bibinfo{author}{\bibfnamefont{J.~G.} \bibnamefont{Checkelsky}},
  \bibinfo{author}{\bibfnamefont{N.~P.} \bibnamefont{Ong}},
  \bibnamefont{et~al.}, \bibinfo{journal}{Nature}
  \textbf{\bibinfo{volume}{460}}, \bibinfo{pages}{1101} (\bibinfo{year}{2009}).

\bibitem[{\citenamefont{Chen et~al.}(2009)\citenamefont{Chen, Analytis, Chu,
  Liu, Mo, Qi, Zhang, Lu, Dai, Fang et~al.}}]{Chen:2009ta}
\bibinfo{author}{\bibfnamefont{Y.~L.} \bibnamefont{Chen}},
  \bibinfo{author}{\bibfnamefont{J.~G.} \bibnamefont{Analytis}},
  \bibinfo{author}{\bibfnamefont{J.~H.} \bibnamefont{Chu}},
  \bibinfo{author}{\bibfnamefont{Z.~K.} \bibnamefont{Liu}},
  \bibinfo{author}{\bibfnamefont{S.~K.} \bibnamefont{Mo}},
  \bibinfo{author}{\bibfnamefont{X.~L.} \bibnamefont{Qi}},
  \bibinfo{author}{\bibfnamefont{H.~J.} \bibnamefont{Zhang}},
  \bibinfo{author}{\bibfnamefont{D.~H.} \bibnamefont{Lu}},
  \bibinfo{author}{\bibfnamefont{X.}~\bibnamefont{Dai}},
  \bibinfo{author}{\bibfnamefont{Z.}~\bibnamefont{Fang}}, \bibnamefont{et~al.},
  \bibinfo{journal}{Science} \textbf{\bibinfo{volume}{325}},
  \bibinfo{pages}{178} (\bibinfo{year}{2009}).

\bibitem[{\citenamefont{Zhang et~al.}(2010{\natexlab{b}})\citenamefont{Zhang,
  He, Chang, Song, Wang, Chen, Jia, Fang, Dai, Shan et~al.}}]{Zhang:2010bu}
\bibinfo{author}{\bibfnamefont{Y.}~\bibnamefont{Zhang}},
  \bibinfo{author}{\bibfnamefont{K.}~\bibnamefont{He}},
  \bibinfo{author}{\bibfnamefont{C.-Z.} \bibnamefont{Chang}},
  \bibinfo{author}{\bibfnamefont{C.-L.} \bibnamefont{Song}},
  \bibinfo{author}{\bibfnamefont{L.-L.} \bibnamefont{Wang}},
  \bibinfo{author}{\bibfnamefont{X.}~\bibnamefont{Chen}},
  \bibinfo{author}{\bibfnamefont{J.-F.} \bibnamefont{Jia}},
  \bibinfo{author}{\bibfnamefont{Z.}~\bibnamefont{Fang}},
  \bibinfo{author}{\bibfnamefont{X.}~\bibnamefont{Dai}},
  \bibinfo{author}{\bibfnamefont{W.-Y.} \bibnamefont{Shan}},
  \bibnamefont{et~al.}, \bibinfo{journal}{Nat Phys}
  \textbf{\bibinfo{volume}{6}}, \bibinfo{pages}{584}
  (\bibinfo{year}{2010}{\natexlab{b}}).

\end{thebibliography}

\end{document}